\documentclass[a4paper, amsfonts, amssymb, amsmath, reprint, showkeys, nofootinbib, twoside]{revtex4-1}
\usepackage[english]{babel}
\usepackage[utf8]{inputenc}
\usepackage[colorinlistoftodos, color=green!40, prependcaption]{todonotes}
\usepackage{amsthm}
\usepackage{mathtools}
\usepackage{physics}
\usepackage{xcolor}
\usepackage{graphicx}
\usepackage[left=23mm,right=13mm,top=35mm,columnsep=15pt]{geometry} 
\usepackage{adjustbox}
\usepackage{placeins}
\usepackage[T1]{fontenc}
\usepackage{lipsum}
\usepackage{csquotes}
\usepackage[pdftex, pdftitle={Article}, pdfauthor={Author}]{hyperref} 
\begin{document}

\title{Extending the range of sizes of monodisperse core-shell hydrogel capsules from composite jet breakup by combined electrical and mechanical actuation}

\author{Lucas Suire}
    \affiliation{LP2N, Laboratoire Photonique Numéerique et Nanosciences, Univ. Bordeaux, F-33400 Talence, France}
    \affiliation{Institut d'Optique Graduate School $\&$ CNRS UMR 5298, F-33400 Talence, France}

\author{Anirban Jana}
    \affiliation{LP2N, Laboratoire Photonique Numéerique et Nanosciences, Univ. Bordeaux, F-33400 Talence, France}
    \affiliation{Institut d'Optique Graduate School $\&$ CNRS UMR 5298, F-33400 Talence, France}
    \affiliation{Treefrog Therapeutics, Pessac, France}

\author{Pierre Nassoy}
    \email[Correspondence email address: ]{pierre.nassoy@u-bordeaux.fr}
    \affiliation{LP2N, Laboratoire Photonique Numéerique et Nanosciences, Univ. Bordeaux, F-33400 Talence, France}
    \affiliation{Institut d'Optique Graduate School $\&$ CNRS UMR 5298, F-33400 Talence, France}

\author{Amaury Badon}
    \email[Correspondence email address: ]{amaury.badon@cnrs.fr}
    \affiliation{LP2N, Laboratoire Photonique Numéerique et Nanosciences, Univ. Bordeaux, F-33400 Talence, France}
    \affiliation{Institut d'Optique Graduate School $\&$ CNRS UMR 5298, F-33400 Talence, France}
    
\date{\today} 

\begin{abstract}
The production of monodisperse particles or droplets is a longstanding issue across various fields, from aerosol science to inkjet printing. In bioengineering, sub-millimeter cell-laden hydrogel capsules have proven valuable for developing in vitro tissue models. A common practical approach for producing such droplets relies on the Plateau–Rayleigh instability to break up a liquid compound jet in air. However, while the droplet size is closely linked to nozzle dimensions, achieving high monodispersity suitable for quantitative biological assays remains challenging due to coalescence events associated with the beads-on-a-string morphology of viscoelastic jets. Here, a microfluidic strategy is introduced, combining electrical and mechanical actuation to enhance control and versatility over jet breakup. By fine-tuning the excitation frequency to select specific modes and applying an electric potential to regulate coalescence, a phase diagram is established, enabling the generation of monodisperse droplets over a broad size range. Notably, a previously hidden effect of the electric field on jet behavior is uncovered and quantitatively characterized. Finally, after crosslinking the compound droplets, capsules with a hydrogel envelope and a core composed of a cell suspension are formed in conditions compatible with cell proliferation, which lay the groundwork for quantitative high precision biological assays  
\end{abstract}

\maketitle







\section{Introduction}
Highly monodisperse liquid droplets are extensively used in a wide range of applications including engine technology, manufacturing, irrigation, ink-jet printing, and bioengineering \cite{eggers2008physics}. For instance, in the case of printing technologies, it is clear that the drop size distribution has a direct impact on the quality and resolution of the final image or product. A multitude of techniques have been used depending on the targeted size of the drops. At low flow rates or, more precisely, small Weber numbers, We=$\rho V_0^2 D/ \sigma$ $<<$ 1, a dripping faucet enables the formation of millimetric droplets, whose size is independent of the diameter of the jet, D,  and solely governed by the interplay between surface tension $\sigma$ and gravity, here represented by the density $\rho$ \cite{de2002gouttes}. Upon increasing We, after a regime of chaotic dripping \cite{shaw1984dripping}, a jetting regime is reached. However, the Plateau–Rayleigh instability makes this jet unstable \cite{plateau1873experimental}. As a result, it fragments into droplets, whose radius is of the order of the diameter of the jet D \cite{rayleigh1879capillary}. In the nanometric-to-sub millimetric range, a lot of variants have been proposed. For instance, nanometric polymeric particles have been produced using a thermally-induced in-fiber Plateau-Rayleigh instability \cite{kaufman2013fiber}. However, at these nanoscales, thermal fluctuations start to play a significant role in the fragmentation \cite{barker2023fluctuating}. Other methods, including ultrasonic nebulizers \cite{kooij2019size}, condensation spray generators \cite{okajima2023investigation}, were developed to generate micrometric aerosols with an intrinsic high polydispersity. With the advent of multiphase flows in microfluidics \cite{nunes2013dripping}, hydrodynamic co-flow-focusing devices enable to generate fairly monodisperse droplets in an immiscible liquid (e.g. water-in-oil) \cite{utada2005monodisperse}, which can further be improved through machine learning-assisted design \cite{lashkaripour2021machine}. In the air, jet breakup is intrinsically irregular, leading to a broad distribution of droplets. When the jet is subjected to a strong electric field, a Taylor cone might be obtained, leading to an electrospraying process that generates fine (typically $<$10 µm) droplets by disintegration of water \cite{taylor1964disintegration}. To prevent coulombic fission when the electric forces exceed the surface tension of the liquid, strategies have been recently implemented by introducing a unipolar ion source that neutralizes the electrospray \cite{carrasco2022monodisperse}. These developments fall in the very active field of electrospray technology. In aerosol science in general, voltage-free alternative techniques were often currently preferred. The first one, invented in the 70s, is based on a piezoelectric-driven vibrating orifice to achieve jet breakup and aerosol generation \cite{berglund1973generation}. The second one, referred to as a pneumatic atomization technique based on aerodynamic flow-focusing, consists in co-flowing a controlled laminar gas stream. By tuning the gas velocity, a narrower distribution of sizes is achieved \cite{ganan1998generation}. In all these cases, the size of the droplets is closely related to the size of the nozzle, and subsequently to the diameter of the jet \cite{rayleigh1879capillary}. Recently, by combining aerodynamic flow-focusing with the application of a periodic mechanical excitation, it was shown that a larger range of monodisperse particles can be achieved with the same nozzle \cite{duan2016generation}.

Altogether, a plethora of techniques from distinct communities has been devised to gain a better control on the Plateau–Rayleigh instability. However, the process of breakup also depends on the nature of the fluid. For instance, when viscoelastic liquids are considered, a beads-on-string structure, where spherical drops are connected by a thin thread, is first triggered before a breakage of the fluid bridges \cite{deblais2018pearling}. This pearling instability often appears to be detrimental to monodispersity since the liquid bridges on each side of the droplets alter the velocity of the droplet through capillary effects and favor coalescence of successive droplets. Very often, for biotechnology applications, biocompatible polymers are used, and hydrogel particles are designed to enable controlled release of active species in drug delivery. More specifically, in tissue engineering, the trend is to encapsulate cells in core-shell permeable particles to favor i) the diffusion of nutrients and waste, and ii) cell-cell interactions within these micro-compartments to avoid anoikis, ie cell death due to the absence of communications with cells or adhesion to the extra-cellular matrix . The prevalent method consists in producing hydrogel (calcium-bridged alginate, photo-reactive gelatin, chemically-reactive polyethylene glycol) core-shells spheres in which the cells are entrapped by co-axial flow in a microfluidic device \cite{kim2011generation,agarwal2013one,yu2015core,fattahi2021core,wang2019one}. However, in all these cases, the capsules formed are surrounded with oil, which may degrade cell survival. An alternative proposed by some of the authors \cite{alessandri2013cellular} and other groups\cite{visser2018air} consisted in performing aerodynamic Rayleigh breakup of a composite jet exiting a co-extrusion device. The sheath of the jet is often composed of alginate and the core is a suspension of cells. Remarkably, after fragmentation into compound droplets, the calcium-induced gelation is so fast that the impact onto the surface of a calcium bath keeps intact the core-shell structure. While the advantage of this method is the increase in the production rate of the capsule because of reduced resistance in the air compared to oil, the distribution of capsules sizes is much broader and bimodal \cite{alessandri2013cellular}, resulting from i) disturbances in the ambient air leading to a poor selection of the fastest growing mode of the instability, and ii) coalescence of successive composite droplets that travel with different speeds in the train of droplets. To perform high-throughput drug screening assays and standardize the analysis, it is crucial to improve the monodispersity of cell-laden capsules in conditions compatible with cell culture and optimized for cell viability. 

To address this challenge, we got inspired by some of the above-mentioned approaches developed in aerosol technology, and investigated how they can be transferred to the practical case of sub-millimetric cellular capsules while taking into account the added complexity of the nature of the gel-forming polymer and the living content. Here, we propose an approach that combines mechanical and electrical actuation to achieve controlled production of compound droplets over a large size range. By applying different actuation frequencies, we can vary the droplet size, and the additional use of electric potential allows us to extend the range of droplet sizes beyond what could be achieved using either technique alone. Specifically, using an injector with a radius R of 95 µm, we are able to produce monodisperse droplets ranging from approximately 170 to 260 µm in radius by tuning the frequency between 800 to 2500 Hz and a DC electric potential between 0 and 6kV. More specifically, we emphasize the role of electric field, in combination with mechanical actuation, in disrupting the beads-on-a-string structure, reducing coalescence and narrowing the size distribution. Finally, we apply this technique with dual electro-mechanical control to the formation of alginate gel-based core/shell spheres and the encapsulation of cells, highlighting the potential of our approach in biomedical applications such as tissue engineering. 

\section{Results}

\subsubsection{
Operating principle of the encapsulation process 
}

\begin{figure*}
  \includegraphics[width=0.85\linewidth]{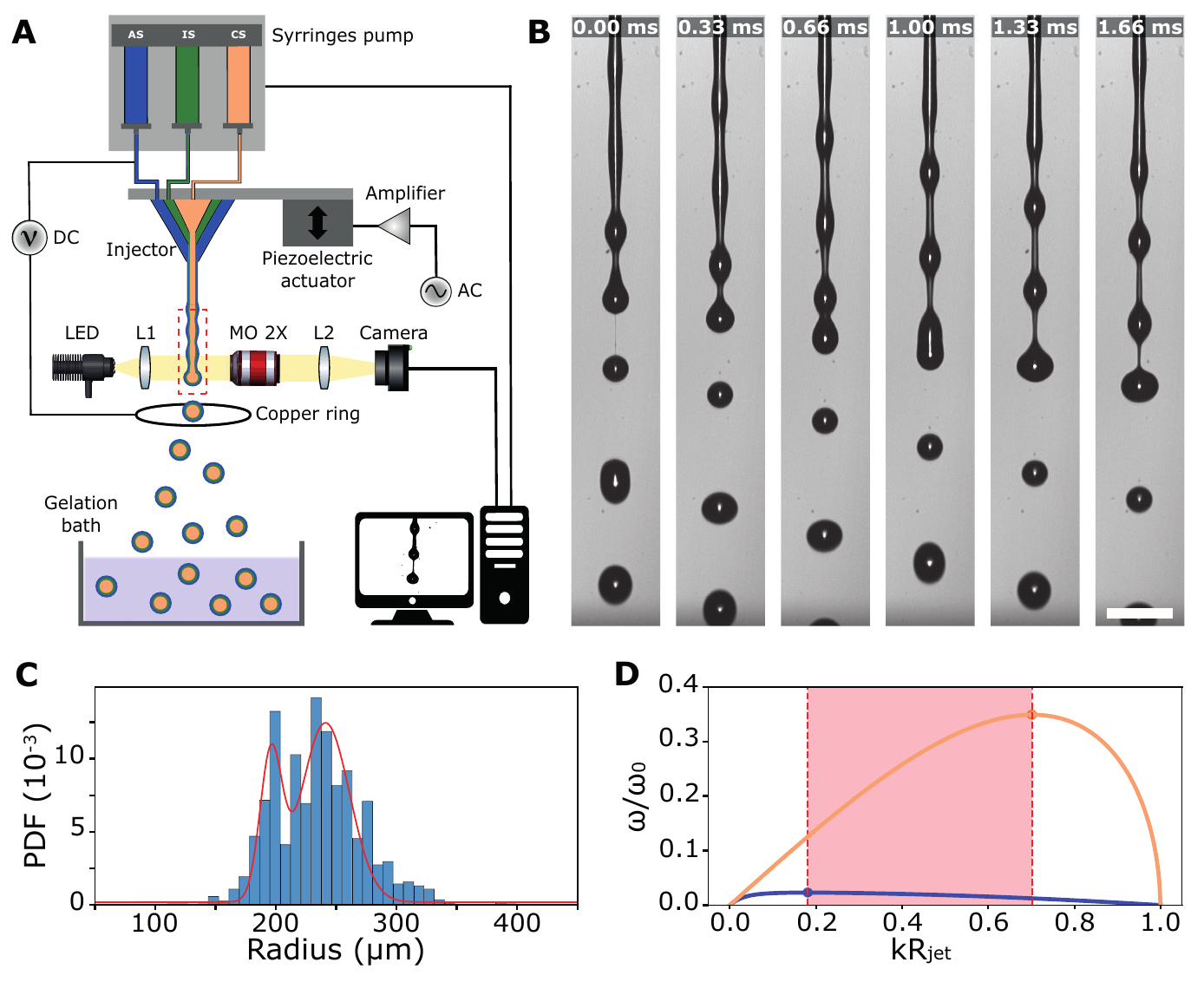}
  \caption{\textbf{Operating principle of the encapsulation platform and high-speed imaging system.} (A) Schematic of the experimental system composed of a 3-way injection system, a co-extrusion injector positioned on a piezoelectric actuator, a ring-shaped electrode in copper and a calcium gelation bath. The three input solutions, cell-suspension (CS), intermediate solution (IS) and alginate solution (AS) are combined in the injector to produce a compound jet which breaks into droplets that are observed with a high-speed imaging system in transmission. (B) Snapshots  showing the fragmentation of the compound jet and a coalescence events. Scale bar, 1 mm. (C) Probability density function of the drop radius evaluated from N=1625 detected drops. The red solid line is an adjusted bimodal distribution with average radii of 196 µm ±9 µm (SD) and 241 µm ±20 µm (SD) respectively. (D) Dimensionless growth rate of sinusoidal perturbation on a cylinder as a function of the dimensionless wave number for pure sorbitol (in orange) and pure alginate (in blue). $\omega_0$ corresponds to the inverse of the capillary time $t_{cap}=\sqrt{\rho R^3 / \gamma}$. Our system acts as a compound jet comprised in the red area.}
  \label{fig1}
\end{figure*}

The experimental set-up used here is adapted from the previously developed Cellular Capsule Technology \cite{alessandri2013cellular}. The microfluidic co-extrusion part remains unchanged, but the actuation and observation parts were drastically modified. The experimental procedure is detailed in the Experimental Section and sketched in Figure \ref{fig1}.A. In brief, the principle consists in using a microfluidic co-extrusion device to produce core-shell capsules made of alginate gel shells enclosing an inner solution (eventually a suspension containing cells). First, a 3D printed co-axial injector combines three fluid flows driven by 3-syringe pumps: the core solution or cell suspension (CS) is injected into the innermost capillary; an alginate solution (AS) flows into the outermost capillary; an intermediate sorbitol solution (IS) flows in the intermediate capillary and serves as a barrier to the potential diffusion of calcium released from cells that, upon gelation, could clog the device. When the flow rate is sufficiently large (We $>$ 1), the fluids emerge from the nozzle as a compound jet, which then breaks up into droplets due to the Plateau-Rayleigh instability. Droplets are produced with a radius that is of the order of the injector diameter, 2R, equivalent to the jet diameter at first order \cite{tomotika1935instability}.  After a fall of approximately 10 cm, droplets enter a gelling bath where the outer alginate shell undergoes gelation mediated by the calcium ions of the bath. Physically crosslinked, transparent and porous capsules are thus obtained. Under the conditions used in this work, unless otherwise specified, we selected R = 95 µm and a total flow rate Q = 165 mL/h, corresponding approximately to 5000 capsules produced per second.	

To capture the dynamics of the drop formation with good temporal resolution, we used a fast imaging system that was positioned around the jet breakup position to observe an area of 9 x 1 $\text{mm}^2$ around the region of interest (Figure \ref{fig1}.B). This bright field imaging system is comprised of a white LED, a collimation lens, a large working distance microscope objective and a tube lens that focuses light on the sensor of a high-speed camera. Acquisitions are typically performed at 12 000 frames per second. As seen on Figure \ref{fig1}.B, such an acquisition speed allows us to dissect dynamics of the Plateau-Rayleigh instability and coalescence events. Qualitatively, we observe that drops are connected by liquid filaments in agreement with the beads-on-a-string structure for viscoelastic liquids such as the alginate solution \cite{goldin1969breakup}. On the sole view of these snapshots, we anticipate that the size distribution of the droplets, and therefore of the capsules, is likely to be very broad. More quantitatively, Figure 1.C displays a histogram of the droplet radii extracted from $>$1000 frames under conventional experimental conditions used previously to generate compartmentalized multicellular assemblies (flow Q = 45/45/75 mL/h for CS, IS and AG respectively). Technical details on the image analysis pipeline are provided in the Experimental Section and in Supporting Information (see Figure S1-S2). A broad and bimodal distribution is observed and can be fitted with two gaussian functions centered around radii equal to $R_0$=196 and $R_1$=241 µm and with respective half-widths 9 and 20 µm. These values are consistent with a main instability mode for a jet of radius $R_0$  and the coalescence between two droplets, doubling the drop volume $R_1$=$2^{1/3} R_0$. However, in our case, a theoretical prediction of $R_0$ is out of reach. The well-described case is the one of an ideal inviscid, pure and non-charged fluid. Our system is more complex since it is compound (alginate shell and water core) and viscous. Previous reports have investigated separately how the dispersion curve is modified when the viscosity varies \cite{eggers2008physics}, when the jet is electrified \cite{hartman2000jet,huebner1971instability,schneider1967stability}. When all these parameters need to be taken into account simultaneously, the complexity of a theoretical description becomes too high or useless for exploring experimentally practical cases, as already pointed out \cite{li2009axisymmetric}. To find the range of radii in which $R_0$ is expected to be found, we used the modified dispersion relation derived by Weber \cite{weber1931zerfall} that gives the growth rate $\omega$ of the instability as a function of the wave number $k$ for viscous fluids. We numerically solved it  for the two pure cases of our composite jet, i.e. pure sorbitol ($\rho$=1000\ kg/m3, $\gamma$ = 72 mN/m, viscosity $\mu$= 1 mPa.s)  and pure alginate jets ($\rho$=1000\ kg/m3, $\gamma$ = 59.8 mN/m, viscosity at 0 shear rate $\mu$= 1.04 Pa.s). The resulting curves are displayed in Figure 1D where the maxima are highlighted, corresponding to the main instability modes of pure sorbitol ($R_0$=180\ µm) and pure alginate ($R_0$=272 µm). Qualitatively, these two limits define the range in k over which our composite jet is expected to exhibit breakup.

By implementing electrical and mechanical actuation, we aim to obtain a homogeneous distribution of sizes for one set of parameters (actuation frequency $f$, electric potential $U$) and, by varying the parameters ($f, U$), to have access to a larger range of capsules sizes. We will thus proceed in two steps. First, we will investigate the conditions to increase the monodispersity of compound droplets in the absence of cells (i.e. IS=CS= sorbitol solution – see Figure \ref{fig1}A). Then, by adding cells in the CS, we will check whether cell-laden capsules exhibit similar features as the drops and whether the proposed experimental conditions do not affect cell survival.

\subsection{Mechanical mode selection}

We first performed a piezoelectric actuation to force the Plateau-Rayleigh instability and select a single mode at a time, instead of multiple modes close to the fastest growing mode in the free Plateau–Rayleigh instability. Several configurations have already been reported to obtain a mode selection by imposing temporal modulations, using a piezoelectric device, onto the velocity of outer phase, i.e. the alginate solution in our case, through a flexible membrane \cite{domejean2017controlled,brandenberger1998new}. Here, we decided to directly mount our microfluidic injector on a piezoelectric actuator, requiring minor change to the experimental setup (see Figure \ref{fig1}A). Here, velocity modulations are replaced by mechanical perturbations. The response of our piezoelectric actuator was fully characterized in terms of response to voltage and bandwidth (see Figure S6).  This led us to keep the magnitude of the displacements constant, of the order of 4 µm, and only vary the frequency. 

\begin{figure*}
  \includegraphics[width=0.85\linewidth]{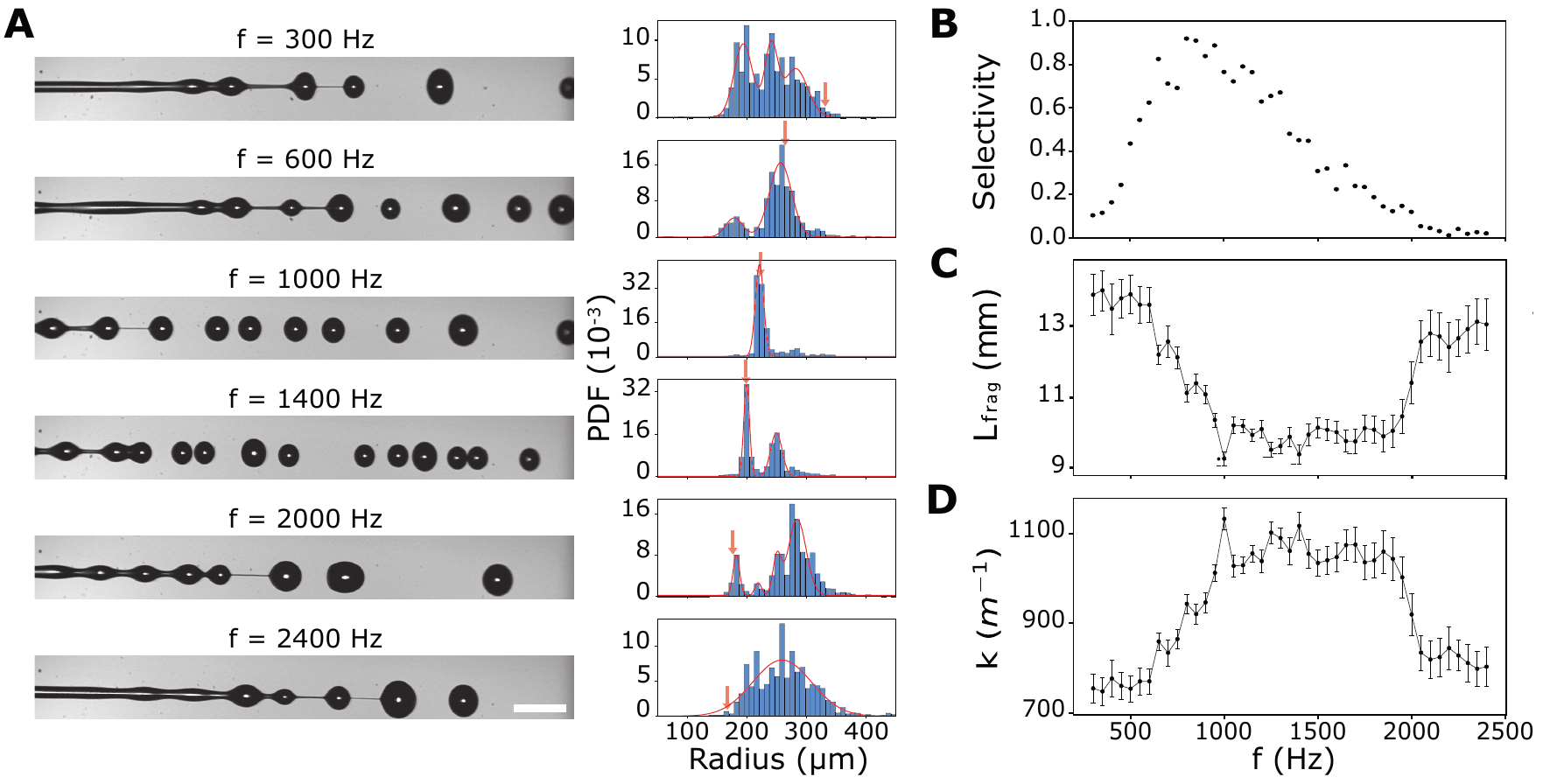}
  \caption{\textbf{Mode selection in Plateau–Rayleigh instability using piezoelectric actuation.} (A) Influence of the piezoelectric modulation frequency $f$ on the jet breakup and corresponding probability density function of the drop radius evaluated for approximately N=2000 detected drops in each case. The flow is from left to right. Red arrows indicate the theoretical drop radius  $R_{th}$, which are not detected at $f$=300 Hz and 2400 Hz, and are found to be decreasing as the frequency is increased: $R_0$=256± 19µm ($f$=600Hz, $R_{th}$=262 µm), $R_0$= 221± 7µm ($f$=1000 Hz, $R_{th}$=221 µm),  $R_0$= 199± 5µm ($f$=1400 Hz, $R_{th}$=198 µm), $R_0$= 182± 6µm ($f$=2000 Hz, $R_{th}$=175 µm). Scale bar, 1 mm. (B) Evaluation of the selectivity parameter $S$  (see definition in the main text) with the frequency $f$ imposed to the piezoelectric actuator. (C) Experimental measurement of the fragmentation length versus the modulation frequency $f$ of the piezoelectric actuator. (D) Experimental dispersion curve showing the spatial growth rate as a function of the modulation frequency $f$ of the piezoelectric actuator.}
  \label{fig2}
\end{figure*}

To understand which modes of the system can be selected, one should refer in principle to the system dispersion relation that describes how the growth rate of the instability varies with the wave number $k$. From the curves displayed in Figure \ref{fig1}D, we anticipate that our system can be forced to a single mode for dimensionless wavenumber  $ k R_{jet}$ ranging from 0.185 to 0.7. With a jet velocity $v_{jet}=Q/\pi R^2$=1.6 m/s, by assuming the jet radius $R_{jet}$ is equal to the injector radius R and using the relation that links the wavenumber to the excitation frequency $f$ of the piezoelectric $k=2 \pi f/v_{jet}$ , we obtain a frequency range from 500 to 1900 Hz approximately, which is below the cutoff frequency of the piezo, as explained in Supporting Information. Experimentally, a slightly larger band is investigated and breakup experiments are conducted for frequencies ranging from 300 to 2500 Hz with a 50 Hz step (Figure \ref{fig2}A). Using volume conservation, we derive that, upon excitation with a frequency $f$, the radius of the droplets is expected to be of the order of:
\begin{equation}
R_{th}= \left( \frac{3 R_{jet}^2 v_{jet}}{4f}\right)^{1/3} 
\label{eq1}
\end{equation}
For low frequencies, typically 300 to 500 Hz, the drop radius presents a very broad distribution due to coalescence events. At this frequency, much lower than the one for which the growth rate is maximal, the amplitude of the piezoelectric is not high enough to force the system. We could extract 3 peaks from the histogram : one at the radius corresponding to the excitation-free instability and two corresponding to the coalescence of two of three droplets. From approximately 500 Hz, we start to observe a mode selection: in the distribution of the radii, a peak is obtained for a radius equals to 256 µm, in agreement with the theoretical values (Equation \ref{eq1}). Yet, other modes, corresponding to smaller satellite droplets due to non-linear effects \cite{chaudhary1980nonlinear} are still present in the system, and the distribution is relatively broad. From 900 to 1200 Hz, effective selection is achieved and the vast majority of droplets have the expected radius within ±10\%.Then, for frequencies above 1200 Hz and up to 2100 Hz, the targeted mode remains mechanically selected but an increasing fraction of  droplets have radii that deviate from the targeted one. Indeed, in this frequency range, the growth rate is expected to decrease. However,  the reduction in droplet sizes and in the distance between two consecutive drops  makes coalescence more likely. Finally, for frequencies higher than 2100 Hz, no mode selection is detected, and we obtain results similar to those obtained for low frequencies. 
For a more quantitative analysis, we characterize the selectivity $S$ , which is defined as:
\begin{equation}
S \left( R_{drop} \right) = \frac{N \in \left[R_{th}\ \pm  10\% \right]}
{N_{total}}
\label{eq2}
\end{equation}
with  $N_{total}$ the total number of drops detected and $R_{th}$  the theoretical drop radius given by Equation \ref{eq1}. As seen in Figure \ref{fig2}B, the curve $S \left( R_{drop} \right)-f$ shows a maximum of selectivity around 1000 Hz with almost 90\% of capsules being in the desired range of radius. A selectivity higher than 75\% is reached for frequencies between 650 and 1150 Hz, corresponding to drop radii between 212 and 256 µm. Away from this range, the selectivity quickly drops for reasons explained before, that are coalescence or small growing rate. From the same experimental data, the fragmentation length $L_{frag}$ is first measured for each frequency (see Figure \ref{fig2}C and Figure S1). The experimental curve shows a significant reduction of $L_{frag}$ for frequencies between 1000 and 2000 Hz, indicating that despite a low selectivity, actuation with the piezo has an effect on the jet instability. From the determination of the fragmentation length and within the linear approximation, the spatial growth rate $k$ scales as 1/$L_{frag}$ and a dispersion curve can be derived (see Supporting Information and \cite{gonzalez2009measurement}). Despite relatively strong assumptions addressed later in the discussion section, the experimental curve displayed in Figure \ref{fig2}D shows a relatively broad plateau with a high growth rate value, indicating that mode selection should be achieved over an extended frequency regime. Notice that this plateau is centered around the frequency of 1500 Hz, for which droplets with a radius of 193 µm are obtained. In the absence of piezoelectric actuation or efficient mode selection, the same result is observed, meaning we retrieve the dominant mode of our system, which corresponds to the highest growth rate. According to Equation \ref{eq1}, $R_{th}$ scales as $f^{-1/3}$, suggesting that the drop radius is eventually weakly affected by the selected frequency. In the next section, we will show how the combination of a variable electric potential and mechanical actuation can extend this range even further.

\subsection{Effect of the electric field on the forced Plateau–Rayleigh instability}

\begin{figure*}
  \includegraphics[width=0.85\linewidth]{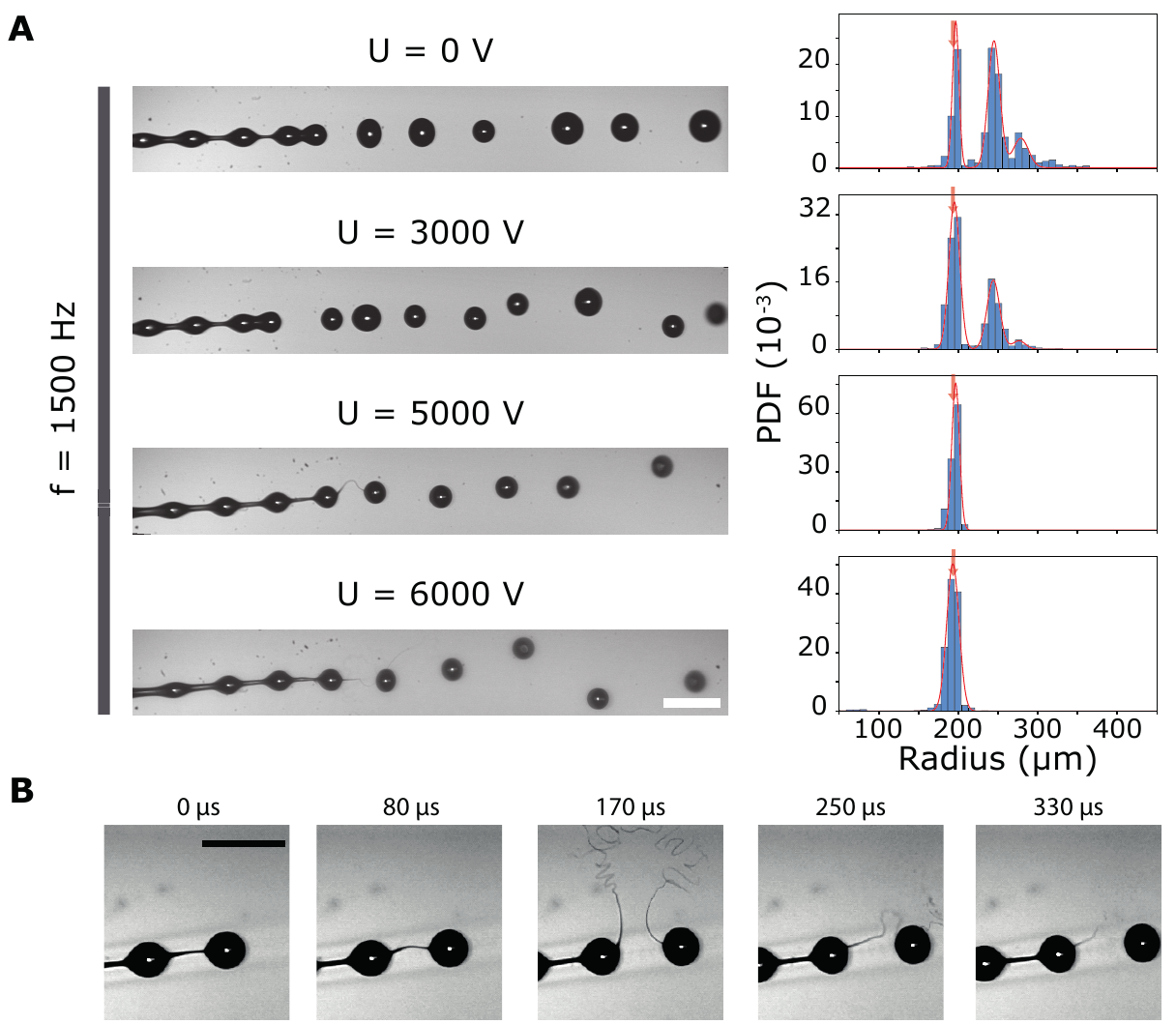}
  \caption{\textbf{Coalescence reduction in the beads-on-a-string regime with a charged jet.}(A) Effect of the electric potential $U$ on the jet breakup and corresponding probability density function of the drop radius evaluated for approximately N=2000 detected drops in each case. The flow is from left to right and the modulation frequency is set to $f$=1500 Hz. Red arrows indicate the targeted radius. Red solid lines are adjusted normal or bimodal distributions. Scale bar, 1 mm. (B) Close up view of a time sequence showing the bending of the viscous filament and its fragmentation at high voltage. Scale bar, 1 mm.}
  \label{fig3}
\end{figure*}

We now investigate the impact of an electric potential onto the jet break-up and the resulting drop radii distribution. First, we set a specific frequency for the piezoelectric and we vary the electric potential $U$ applied between the jet and a metallic copper ring positioned near the fragmentation length. We selected a frequency $f$ equals to 1500 Hz, for which the selectivity is good ($S \sim 0.5$) but not maximal. As seen on the bright field images in Figure \ref{fig3}A, mode selection without any electric field is partly efficient as drops exhibit a bimodal distribution. The peak at larger radii reflects the merging of two drops from the main mode. Note that a third peak resulting from the coalescence of 3 droplets can be detected around $R_2=3^{1/3} R_0$ with a lower magnitude associated to a lower probability of two successive coalescence events. 

When the electric potential is increased to 3 kV, the ratio between the sub-populations changes: the number of coalesced drops decreases and the targeted population becomes the dominant one. If the electric potential is further increased to 5 kV, the histogram shows that only drops whose radius is equal to $R_{th} \pm 5 \%$ are produced. For such a high electric potential, even if the dispersion ring is well below the fragmentation length ($\sim $7 mm), drops are slightly deviated from the initial axial position and bending of the filament appears. Both of these effects are even more noticeable when the electric potential reaches 6 kV. We are getting closer to a whipping regime \cite{guerrero2014whipping}, that we will discuss more in details below.  This sets an upper limit of the electric potential as it starts to destabilize the core/shell structure of the jet and drops, thus preventing the formation of capsules after hydrogel reticulation (see Figure S5).    

Here, at first sight, we confirm that the effect of charging the alginate tends to reduce the coalescence and improve the monodispersity of the droplets. One obvious and previously reported explanation \cite{brandenberger1999monodisperse} is that the coulombic repulsion causes them to self-disperse. However, a closer look reveals that above a given electric potential ($\sim $2500 V), the viscous filament that connects two consecutive drops exhibits deformations reminiscent of buckling before fragmenting into a complex pattern (Figure \ref{fig3}B). This early break-up of the filament suppresses the unbalanced capillary force that brings the two droplets towards merging. These experiments demonstrate the benefit of increasing the electric potential for a given frequency applied by the piezoelectric actuator. In the next section, we will investigate the effect of this potential as a function of the frequency, and how to define a pair ($f,U$) to obtain a good selectivity over a wider range.

\subsection{Combination of electric field and mechanical actuation for tunable monodisperse capsule production}

\begin{figure*}
  \includegraphics[width=0.85\linewidth]{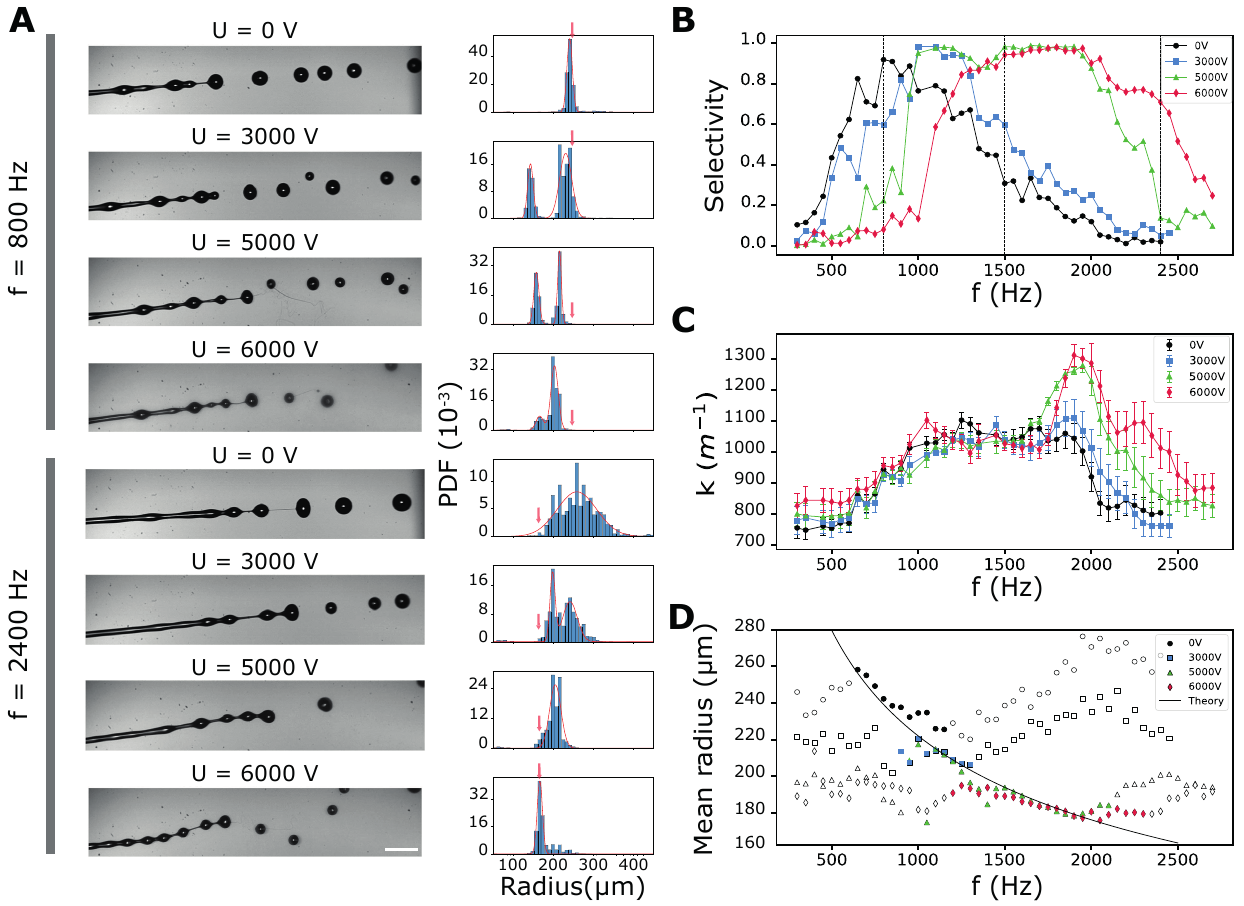}
  \caption{\textbf{Tunable and homogeneous drop production with mechanical actuation and varying electric potential.} (A) Effect of the electric potential $U$ on the jet breakup for two specific modulation frequencies $f$=800 Hz and $f$=2400 Hz. The flow is from left to right. Scale bar, 500 µm. (B) Evaluation of the selectivity $S$ versus the modulation frequency $f$ for several values of electric potential $U$. (C) Experimental dispersion curves of our system that displays the growth rate versus the modulation frequency $f$ for several values of the electric potential $U$. (D)  Evolution of the drop radius as a function of the modulation frequency $f$. Solid symbols indicate conditions where the selectivity is higher than 75\%. Black solid line corresponds to the theoretical model given by Equation \ref{eq1} with no adjustable parameters.}
  \label{fig4}
\end{figure*}

Following the proof-of-concept experiment conducted in the previous section, we now investigate the combined effects of the electric potential $U$ and the actuation frequencies $f$ over the following ranges, $U$: 0 to 6kV; $f$=300 to 2700 Hz. The other experimental parameters are kept constant (solution composition, injector diameter $R$ and flow velocity). The results are summarized in Figure \ref{fig4}B-D. First, we focus on two extreme frequencies to exemplify the jet behaviors and features of the droplet distributions: $f_1$=800Hz is taken at the edge of the range of frequencies for optimal selectivity, and $f_2$=2400 Hz far above the optimal frequency without electric potential. For $f_1$=800 Hz, the satellite droplets that were absorbed by the primary drops at $U$=0V are now repelled when $U$ is varied (Figure \ref{fig4}A top) which leads to poor selectivity ($S$=0.05 at $U$=6000V). Altogether, this specific case shows that the presence of charges in conjunction with mechanical actuation is not always beneficial to selectivity. 

For $f_2$=2400 Hz, at a frequency where the piezoelectric actuation alone was not able to force the system to produce monodisperse droplets ($S$=0.05 at $U$=0V), the presence of charges helps to obtain a narrower distribution (Figure \ref{fig4}A, bottom). However, the size of the obtained droplets differs from the expected one; we mainly obtain the size associated with the dominant mode without excitation, meaning that coalescence is reduced without mode selection actually being effective. Above 5 kV, the selectivity is significantly improved (Figure \ref{fig4}B) up to approximately 70\% at 6 kV, even if this voltage value corresponds to the onset of the whipping regime. Here we observe the shift of the dispersion curve towards higher frequencies when the jet is electrified.
Hence, depending on the forcing frequency $f$, the monodispersity of the droplets after break-up can be increased either by decreasing or increasing the electric potential $U$. Figure \ref{fig4}B displays the evolution of the selectivity as defined by Equation \ref{eq2} as a function of $f$ for different electric potentials. We observe that it is possible to produce drops of a given size with a selectivity higher than 75\% over a range between 650 and 2300 Hz if $U$ is adapted properly. There is a shift of the selectivity towards the higher frequencies when $U$ is increased. This is also consistent with the experimental dispersion curve of our hydrodynamic system. From the measure of the fragmentation length $L_{frag}$, the growth rate is estimated for each modulation frequency $f$ and each electric potential applied $U$.  A shift of the dispersion curve towards high frequencies is observed when $U$ increases, indicating that, when the jet is electrified, new modes become excitable with a mechanical actuation. This is qualitatively in agreement with theoretical predictions \cite{hohman2001electrospinning}. From a practical point of view, a representation of the effective drop size as a function of $f$ can be more meaningful. Figure \ref{fig4}D provides such a representation accompanied by the theoretical curve corresponding to Equation \ref{eq1}. Solid symbols indicate conditions where selectivity exceeds 75\%. A production of drops of radius between 170 and 260 µm with an excellent monodispersity is within reach under the synergistic combination of electrical and mechanical stimulation with the same nozzle.

\subsection{Homogeneous production of capsules for 3D cell culture}

\begin{figure*}
  \includegraphics[width=0.85\linewidth]{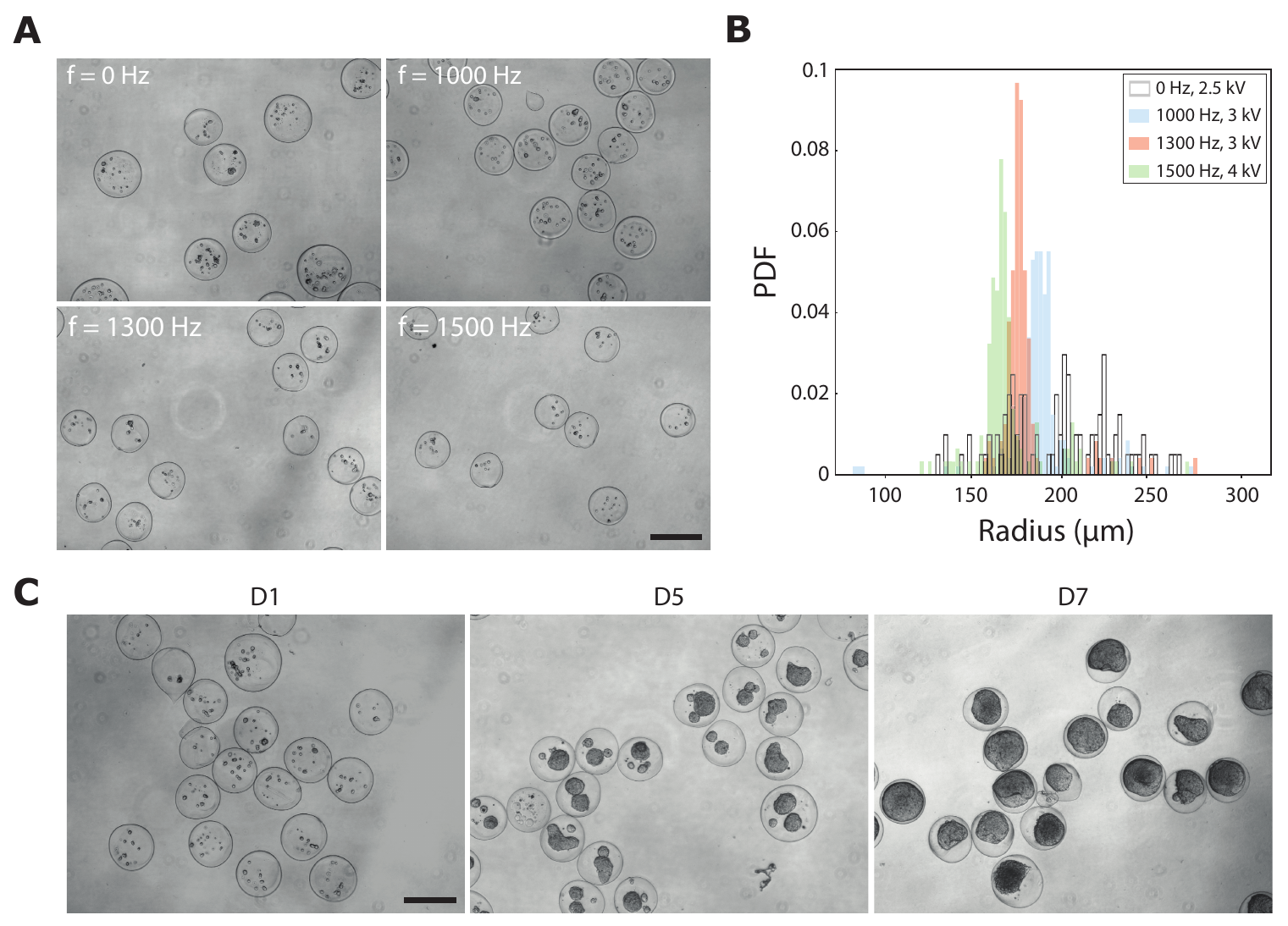}
  \caption{\textbf{Formation and growth of HEK 293T cell spheroids in hydrogel capsules.} (A) Bright field images of capsules at day 1 after encapsulation for modulation frequencies equal to 0, 1000, 1300 and 1500 Hz. Scale bar, 500 µm. (B) Probability density function of the capsules radius evaluated for N=50 capsules for each condition. (C) Bright field images of capsules for modulation frequency $f$=1300 Hz at days 3,5 and 7 after encapsulation. Scale bar, 500 µm.}
  \label{fig5}
\end{figure*}

Our initial goal was to develop a robust method to produce well-calibrated capsules for 3D cell culture. The drawback of the previously developed technology based on a coextrusion device without actuation \cite{alessandri2013cellular} is that the mean size of the (broad) distribution of capsules is uniquely given by the diameter of the nozzle. The question addressed here is to generate monodisperse capsules  and with adjustable radius without having to change the nozzle. We used the modified version of the Cellular Capsules Technology described in Figure \ref{fig1} and replace the sorbitol solution in the central channel (CS) with  a cell suspension that contains human embryonic kidney cells (HEK 293T). To assess the performances of the above-mentioned approach, we generated cellular capsules in four different configurations, without any mechanical actuation ($f$=0 Hz) and for excitation frequencies equal to 1000, 1300 and 1500 Hz. For all these configurations, the electric potential is set to obtain maximum selectivity according to the  findings in the previous section, ie  2.5, 3, 3 and 4kV respectively. With a cell concentration of 500 000 cells/mL, assuming a Poisson distribution in the capsules of average radius 200 µm, we expect 3 to 6 cells encapsulated per capsule. Once the compound droplets have fallen in the gelation bath, we first observed that the shape of the capsule is not altered by the impact and that cells are indeed trapped inside the capsules, as shown previously \cite{alessandri2013cellular,cohen2023engineering}. While capsules obtained at $f$=0 Hz and $U$=2.5 kV, which corresponds to the standard conditions used in previous reports, exhibit a strong polydispersity (Figure \ref{fig5}A left), the distribution of sizes as the actuation frequency increases is significantly affected. Capsules are observed to be smaller,  in line with Equation \ref{eq1},  and monodisperse, provided that $U$ is adjusted according to Figure \ref{fig4}B between 3 and 4 kV. More quantitively, segmentation was performed over more than 50 capsules in each condition, and the resulting distribution of radii is displayed in Figure \ref{fig5}B. The presence of cells does not cause any significant broadening of the distribution. Note however that there is a 15\% difference between the observed size of drop in the air and the measured capsule size in the calcium bath (see Table S2). This contraction of alginate upon gelation is due to osmotic pressure and is well documented \cite{martinsen1989alginate,huang2017size}.
In all these cases, the core-shell structure was preserved (see Figure S5); this aspect is of importance as the available volume for the cells depends both on the outer diameter and the alginate shell thickness. Finally, we monitored the fate of encapsulated cells over several days in standard culture conditions ($f$=1300 Hz, $U$=3kV). Bright field microscopic images of a population of capsules inside which cells proliferate are shown in Figure \ref{fig5}C. From the few cells present at day 1, aggregates rapidly form, grow and merge as spheroids quickly fill the capsule.  This demonstrated the low invasiveness of our method despite the use of a relatively large electric field and its applicability to produce well-calibrated bioreactors for 3D cell culture. 

\section{Discussion}

These past 20 years, 3D cell culture models have emerged as more  physiologically relevant than traditional 2D culture to model diseases and organ functions \cite{cukierman2001taking}. Despite recent efforts and advances, especially in the microfluidics and microfrabication fields \cite{eisenstein2018organoids,zhao2022organoids,hofer2021engineering}, there is still  a need for a standardized method to develop a reproducible and well-controlled production of multi-cellular aggregates for  high-throughput investigations. Here, we proposed a novel method to generate  monodisperse  and size-tunable hydrogel capsules for 3D cell culture. This approach relies on a synergetic mechanical and electrical actuation onto the fragmentation of a compound aqueous jet in air.  With minor modifications to the experimental setup previously developed, we managed to significantly improve the monodispersity and versatility of the technique. Indeed, we   demonstrated that the capsule size could be tuned over a wide range (radius varying between 170 and 260 µm) at high throughput (5000 capsules per seconds) and in oil free conditions, that is in conditions optimal for cell viability. 

We first investigated how a piezo actuation can select a mode, ie the size of a compound drop composed of a sheath of alginate solution and a core of sorbitol, by scanning all excitation frequencies between the maxima in the dispersion relations for the pure solutions of alginate 2\% and sorbitol. We found, as already claimed or reported in the literature \cite{maitrejean2024comprehensive,domejean2017controlled,brandenberger1998new}, that mechanical actuation by itself can indeed produce monodisperse droplets only in a limited range, from 220 to 260 µm in our case (nozzle R=95 µm). The accessible range of monodisperse droplets can be extended either by reducing the coalescence or by increasing the range over which the mode selection operates. In this work, we further demonstrated that a properly selected electric potential combined with a mechanical actuation also helps on both aspects. In particular, we focused on the following three beneficial effects of the electric potential: (1) the reduction of coalescence by coulombic repulsion, (2) the breaking of the viscous filament above a certain threshold,  and (3) the shift of the dispersion relation towards high wavenumbers as a function of the jet charge.

\begin{figure}
  \includegraphics[width=0.95\linewidth]{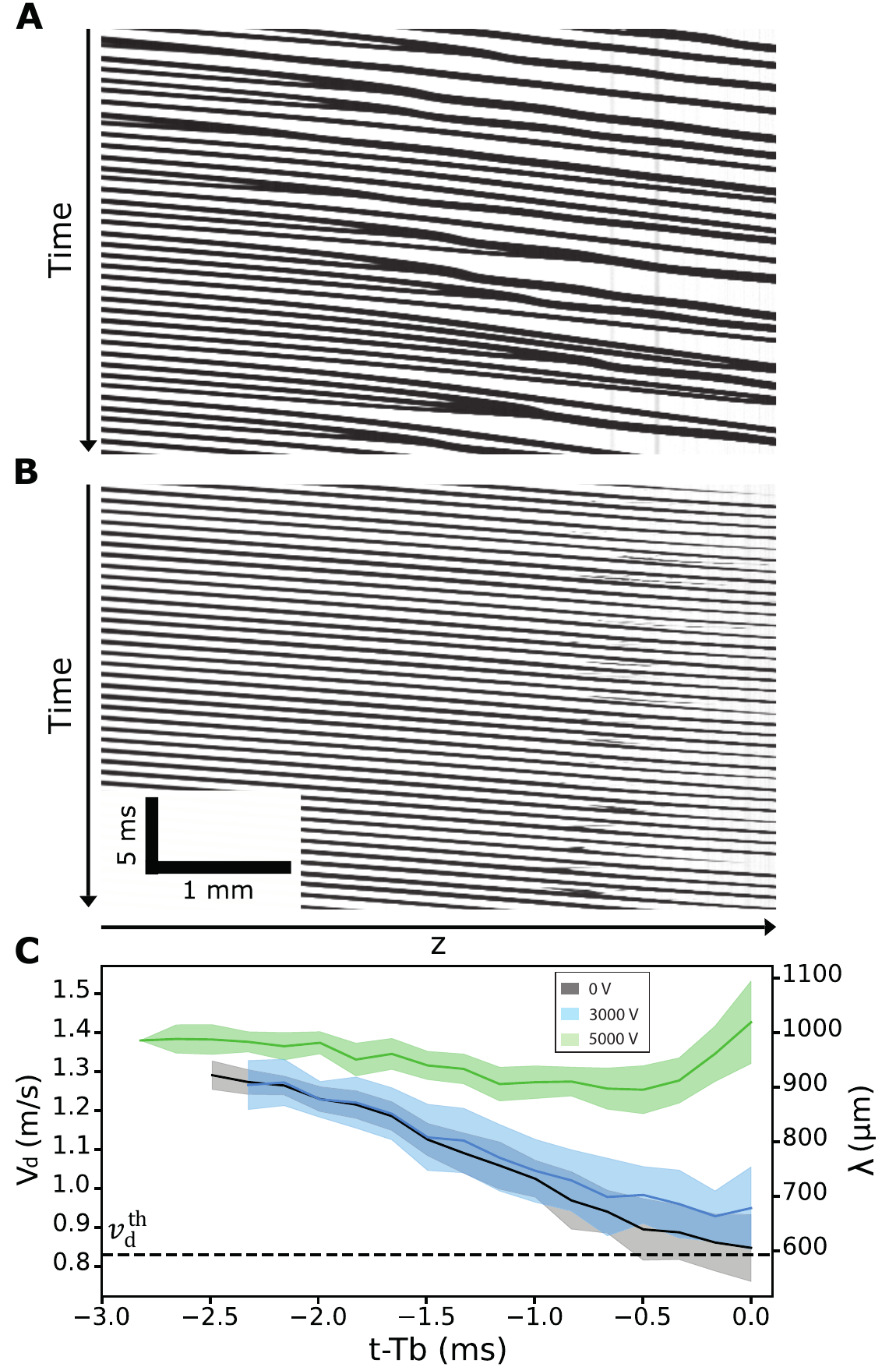}
---  \caption{\textbf{Droplet speed and spacing close to fragmentation.} (A, B) Spatio-temporal diagram of jet fragmentation for an excitation frequency $f$=1400 Hz and an electric potential $U$=0V and $U$=5000V respectively. (C) Evolution of velocity $v_d$ and spacing $\lambda$ as a function of time before break-up, $T_b$. N=16 measurements in each case.}
  \label{fig6}
\end{figure}

Before fragmentation of the jet, droplet coalescence originates from two main causes: the capillary force induced by the thin viscous filament between two consecutive droplets of radius $r_f$, $F_{cap}=2\pi R r_f\gamma$, and the variation in velocities of different droplets within the "beads-on-the-string" regime, primarily due to small fluctuations in pump flow rates. Considering a drop connected by two filaments, the downstream filament has a smaller radius than the upstream one resulting in a capillary force directed towards the injector, causing the drops to slow down until it merges with the following one. By analysing images sequences similar to the one shown in Figure \ref{fig3} and \ref{fig4}, we plotted a kymograph in Figure \ref{fig6}A.  It provides a spatio-temporal representation of the trajectory of droplets inside the beads-on a-string structure before break-up time $T_b$ where each dark line represents a droplet in the beads-on-a-string" regime, meaning that the slope is the associated velocity. In the absence of an electric field, merging lines correspond to coalescence events  that are observed to be numerous, due to the slowing down of the droplets caused by the capillary force of the thread (“string”). However, when an electric potential $U$ = 5000 V is applied between the jet and the metallic ring, the velocity of the drops remains constant, and coalescence events are very scarce for the same excitation frequency $f$=1400 Hz (Figure \ref{fig6}B). 

 While coulombic repulsions have the obvious effect of dispersing the charged droplets after break-up, the effect of the electric potential is more subtle before fragmentation. We measured the instantaneous velocity $v_d$ of droplets as well as their average center-to-center spacing $\lambda$ for the same excitation frequency $f$=1400 Hz and for $U$=0, 3 and 5 kV. Figure \ref{fig6}C shows the evolution of $v_d$ and $\lambda$ as a function of the time before break-up. In the absence of any electric potential, the droplets transition from an initial velocity equal to that of the jet (1.6 m/s) to a velocity of approximately 0.86 m/s (black line, Figure \ref{fig6}C) at the fragmentation ($t-T_b$=0). Note that this value is in excellent agreement with the theoretical velocity derived from mass and momentum conservation and reported by Schneider \cite{schneider1967stability}: 
 \begin{equation}
v_d \sim v_{jet} - \frac{2 \gamma}{ \rho R_{jet} v_{jet}} \sim \ 0.83 \ \text{m/s}
 \end{equation}

This decrease in instantaneous velocity leads to a reduction in the average spacing between two droplets, which reaches a value of approximately 600 µm at fragmentation. Since the capillary length, $L_c=(\gamma / \rho g)^{1/2}$, the length over which the liquid surfaces are deformed is of the order of 1 mm, we expect that droplets separated by a liquid bridge of length  600 µm $<$ $L_c$ undergo a significant attraction force that increases the probability of coalescence before fragmentation. Similar results are obtained for $U$=3 kV (blue line, Figure \ref{fig6}C). However, for $U$=5 kV, a completely different behavior is observed as the jet approaches fragmentation. Following a slower decrease of velocity and spacing, $\lambda$ and $v_d$ exhibit an increase within the last 300 µs prior to fragmentation, preventing the droplets to coalesce. This originates from the interplay between the capillary force, which is controlled by the radius of the filament and dominant in the first phase, and the electrostatic force between droplets charged with an electrical density $\sigma_0 \approx \frac{\epsilon_0 U}{R_{jet}}$ , which dominates the terminal phase (with $\epsilon_0$ the vacuum permittivity).

Our second major observation is related to the direct effect of the electric field on the fragmentation of the compound jet. There is  a threshold value of $U$=2500 V (Figure \ref{fig3}B and Movie S3, Supporting Information), over which the viscous filament connecting the droplets in the "beads-on-a-string" regime undergoes buckling-like phenomenon. We propose this process to be analogous to the whipping instability of a jet \cite{guerrero2014whipping}. The transition between the capillary (Rayleigh) instability and the whipping instability is set when the electric Bond number $\Gamma=\frac{\epsilon_0 U^2}{R_{jet}\gamma}$ is of the order of unity. The whipping instability of the jet occurs at about 7000 V under our experimental conditions. Applying the same reasoning to the filament between consecutive droplets, which is  approximately ten times smaller than the jet itself, we obtain a whipping threshold value of 2500 V for the filament, in agreement with our experimental observations. To our knowledge, this electric effect that suppresses capillary forces responsible for droplet coalescence by triggering whipping of the filament between viscoelastic droplets has never been reported in the literature.

Finally, the last notable effect of the electric field concerns the shift of the dispersion curve toward higher wavenumbers when the electric potential applied to the jet increases (Figure \ref{fig4}C). The main beneficial effect is, of course, the ability to select higher frequencies, leading to the formation of smaller droplets. However, this also modifies the cutoff frequency associated with the second harmonic. Without an electric field, low frequencies allows for the generation of a highly homogeneous droplet population. In contrast, applying the field at the same frequency introduces the presence of two distinct droplet sizes, associated with the piezoelectric excitation frequency and its double frequency.

\begin{figure}
  \includegraphics[width=0.85\linewidth]{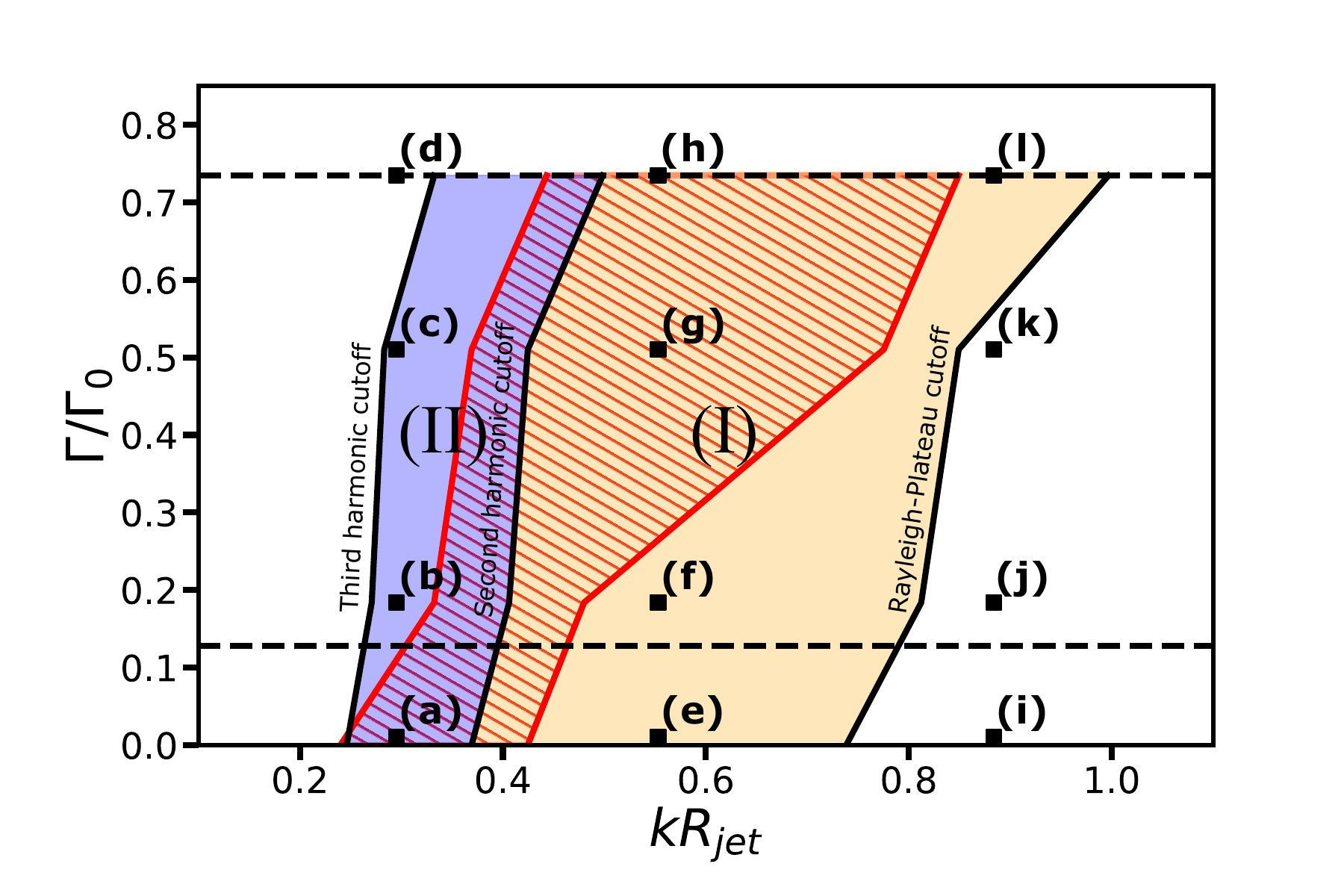}
  \caption{\textbf{Jet response and selectivity mapped in the $k R_{jet}-\Gamma$ diagram}. Region I: Plateau–Rayleigh monomodal instability. Region II: Plateau–Rayleigh bimodal instability. Cutoff wavenumbers are defined as values for which growth rate is equal to 850 $\text{m}^{-1}$, when modes are not selected and the dispersion curve reaches a plateau.  Red hatched region corresponds to  selectivity $S$ $>$ 75\%. Dashed horizontal lines correspond to the whipping threshold of the jet and the filament. Scattered data points are experimental data associated with the following pair of experimental parameters ($f$, $U$): a=(800 Hz, OV), b=(800 Hz, 3000V), c=(800Hz, 5000V), d=(800Hz, 6000V), e=(1500Hz, 0V), f=(1500Hz, 3000V), g=(1500Hz, 5000V), h=(1500Hz, 6000V), i=(2400Hz, 0V), j=(2400Hz, 3000V), k=(2400Hz, 5000V), l=(2400Hz, 6000V). }
  \label{fig7}
\end{figure}

Altogether, these observations that result from  the investigation of jet break-up as a function of potential $U$ and frequency $f$, allow us to define a phase diagram representing the different observed regimes as a function of dimensionless parameters, specifically the electric Bond number and the product of the wavenumber $k$ by the jet radius $R_{jet}$ (see Figure \ref{fig7}) \cite{yang2014crossover}. The boundaries of the different regimes are set by the experimental observation of the cutoff frequency (minimum growth rate) for the Plateau-Rayleigh instability, the $\text{2}^{nd}$ harmonic cutoff and the whipping threshold for both the jet and the viscous filament. On top of this phase-diagram, we superimposed the hatched area between the two red lines where a selectivity higher than 75\% is obtained. This dimensionless representation will assist those who wish to apply our approach to their own system in identifying an operating point that provides the generation of a monodisperse droplet population for a given jet size and, consequently, a specific injector size. 

The previous diagram is expected to guide bioengineers in the tailored production of capsules for 3D cell culture applications. In particular, in this work, we checked that we could produce monodisperse production of capsules in culture conditions (see Experimental Section) with a low seeding density (5 cells in average).  The transition from droplets to capsules however requires  gelation of the alginate shell. Charged droplets falling at high speed into a calcium bath can in principle be damaged or deformed. For instance, jet whipping appears to disrupt the laminar flow, preventing the formation of capsules with a core-shell structure. Above 7 kV, which corresponds to the whipping threshold, we observed that the number of ruptured capsules reached 90\%. These detrimental effects can be significantly reduced by increasing the thickness of the alginate shell by tuning the ratio between the inner (CS and IS) and outer (AS) flow rates \cite{alessandri2013cellular}. Remarkably, increasing the alginate thickness by 1.5-fold completely solved the issue (see Figure S5).

In summary, we have presented and characterized a technique to generate monodisperse hydrogel capsules at high throughput by providing  a better control on the Plateau–Rayleigh and gaining insight in the underlying physical mechanism. Besides being an attractive solution to produce calibrated complex 3D systems, the finding that the combination of electric field and mechanical actuation can also open new directions to study more fundamental physical phenomena such as the coupling of jetting and whipping mode or the production of drops with different sizes at the same time. The variation in satellite droplet size with excitation frequency and voltage, along with the preservation of their core-shell structure, could further extend the range of selectable droplet radii for the same injector. We anticipate that investigating the effect of the electric potential with different geometries or with a temporal modulation can provide further insights into this complex system.

\section{Experimental Section}
\subsection{Encapsulation}
\subsubsection{Microfluidics Injection System}

Three fluids—the shell solution, the intermediate solution, and the cell suspension—are transported through PTFE 560 tubing with an internal diameter of 1 mm (IDEX Health \& Science, Tub PFA 1507L). The flow rates are precisely regulated using a syringe pump (CETONI NEMESYS LOW PRESSURE SYRINGE PUMP, NEM-B101-02 E). These fluids are subsequently combined to form a compound jet via a custom-designed 3-way micro-injector equipped with a nozzle having a radius of 95 µm in the present work. The co-extrusion micro-injector is mounted on a manual stage that facilitates both horizontal and vertical adjustments.

\subsubsection{Electromechanical Control}

The micro-injector is clamped to a piezoelectric actuator (Piezosystem Jena, Model 76579), which oscillates at a frequency that mechanically controls the instability of the jet (see Figure S6A). This actuator is driven by an arbitrary waveform generator (TTi, TGA1242), whose  signal is amplified by an amplifier system unit (Piezosystem Jena, ENV 400). A custom-made metallic copper ring is employed to apply an electric field to the jet. This copper ring is connected to a sliding support, allowing reproducible vertical positioning. An electrode, linked to a DC voltage generator (Standford Research System, PS375 / +20 kV) charges the alginate prior to its entry into the injector, thereby creating an electrical potential difference between the ring and the jet.

\subsubsection{Gelation}

Following the break-up of the jet, compound droplets are formed, with an aqueaous core or a cell suspension and surrounded by an alginate spherical sheath. These drops fall in a calcium bath positioned 10 cm below the injector and on the surface of which a small quantity of surfactant (Tween 80) was added to minimize the impact. Calcium ions mediate instantaneous cross-linking of the outer polymeric part of the droplets into alginate hydrogel capsules. When cells are added to the internal solution, they remain trapped  in the core of the capsules \cite{alessandri2013cellular}.

\subsection{Optical detection}

\subsubsection{Image acquisition of the droplets}

The optical assembly consists of a white light-emitting diode (Thorlabs, MWWHL4) followed by a collimation lens (Thorlabs, AC254-030-A-ML). A long working distance  objective lens (Thorlabs, TL2X-SAP, NA=0.1, WD=56.3 mm) collects the light, which is then focused by a tube lens (Thorlabs, AC508-150-A-ML) onto the sensor of a high-speed camera (Mikrotron EoSens 1.1 CXP2). For a field of view of around 8.5×2.5 $mm^2$, this optical system provides a frame rate of approximately 12,000 frames per second with a spatial resolution of 8.5 µm. The exit of the micro-injector, and thus the beginning of the jet in air, is positioned approximately  6 mm  above the field of view of the camera. Before each acquisition series, the exact position of the injector is calibrated, enabling precise measurements of various parameters including the fragmentation length $L_{frag}$. For each experimental condition, a video with 2000 frames was recorded.

\subsubsection{Image processing of the droplets}

Two following key parameters were extracted from each acquisition : the fragmentation length and the droplet radius. 	
(i) The fragmentation length was estimated using a custom Python script with the Numpy, Matplotlib, OpenCV, Scipy, and Scikit-image libraries. Initially, the script reads the videos and, if necessary, apply a rotation to the image to account for any slight deviation of the jet from the vertical axis (see Figure S1). Then, we performed a binarization of the images to distinguish the jet from the background. The pixels in the central portion of the video, which contain the jet, are then summed along the small axis to obtain an intensity profile as a function of the direction of the jet trajectory. For each frame, the script detects the position of the first zero in the intensity profile, which corresponds to the end of the continuous jet. This position is plotted as a function of time, and the peaks are detected, which allows to derive the fragmentation length values. For a video of 2000 frames, the script detects at least 100 fragmentation length values, which are then averaged to yield a statistically robust final value.	
(ii) The size of the drops was measured using a Python code with the Numpy, Matplotlib, OpenCV, and Scikit-image libraries (see Figure S2). Initially, the code selects a user-defined region of interest (ROI) which spans a length of 1500 µm. This ROI is positioned 1.5 mm below the fragmentation length to ensure sufficient time for the oscillations in drop shape to dampen out. Then, a Canny filter is applied to detect the contours of the drops based on intensity gradients. This approach facilitates the detection of only in-focus droplets. Finally, the drops are detected using a circular Hough transform, which yields their radius.

\subsubsection{Image acquisition and processing of the hydrogel capsules}

Images of the capsules were captured using a camera (Nikon DS-Fi1) mounted on a phase contrast imaging microscope (Nikon Eclipse TS100) with a resolution of 2.32 µm per pixel. A Python script was employed to estimate the radius of the capsules after gelation. This script uses the Numpy, Matplotlib, Scikit-image, and Cellpose libraries. Individual capsules entirely within the image frame are detected and segmented with Cellpose \cite{stringer2021cellpose}. Then, the script measured the volumes of the resulting masks and computed the radius of the corresponding spheres.
All the codes developed for this project are accessible here \cite{gitbiof}.

\subsection{Cell culture and encapsulation}

HEK293T cells (gift from the Bordeaux Institute of Oncology) were cultivated in plastic flasks (Corning, cat. no. 353136) and sustained in Dulbecco’s Modified Eagle Medium (DMEM) (Biowest, cat. no. L0103-500), enriched with 10\% fetal bovine serum (Capricorn, cat. no. FBS-16A) and 5\% penicillin-streptomycin (Gibco, cat. no. 15140122). The cells were kept in an incubator  (water-saturated atmosphere,5\% CO2, 37°C).

\subsection{Chemicals}

The alginate solution was prepared by dissolving 2\% wt/vol sodium alginate (AGI, cat. No. I3G80) in Milli-Q water, followed by filtration at 1 µm and storage at 4°C. The sorbitol solution was obtained by dissolving 300 mM sorbitol (Sigma-Aldrich, cat. No. S1876) in Mili-Q water. The calcium bath was prepared by dissolution of 100 mM calcium chloride (Calcium chloride hexahydrate, 98+ \%, Thermo Scientific No. 389262500). Traces of of Tween 80 (Merck) were added to the surface of the bath just before running the encapsulation process. \\

\textbf{Supporting Information}

Supporting Information is available from the Wiley Online Library or from the author.
\bigskip

\textbf{Acknowledgements}  

The authors thank all the BiOf team for fruitful discussions and especially Adeline Boyreau for cell culture. This work was funded by the French National Agency for Research (ANR-21-CE18-0038, ANR-23-CE45-0016, ANR-22-CE42-0019), Idex Bordeaux (Research Program GPR Light), the EUR Light S\&T (PIA3 Program, ANR-17-EURE-0027) and the PEPR ENVie (ENVie : ANR-24-EXME-0002).
\bigskip

\textbf{Data Availability Statement} 

The data that support the findings of this study are available from the corresponding author upon reasonable request.

%
\bibliographystyle{MSP}
\bibliography{Capsules_Lucas}




\end{document}